# Correlation between COVID-19 Morbidity and Mortality Rates in Japan and Local Population Density, Temperature, and Absolute Humidity

**Sachiko Kodera [1], Essam A. Rashed [1,2] and Akimasa Hirata [1,3,\*]**

[1] Department of Electrical and Mechanical Engineering, Nagoya Institute of Technology, Nagoya 466-8555, Japan; kodera@nitech.ac.jp (S.K.); essam.rashed@nitech.ac.jp (E.A.R.)
[2] Department of Mathematics, Faculty of Science, Suez Canal University, Ismailia 41522, Egypt
[3] Center of Biomedical Physics and Information Technology, Nagoya Institute of Technology, Nagoya 466-8555, Japan
\* Correspondence: ahirata@nitech.ac.jp; Tel.: +81-52-735-7916

**Abstract:** This study analyzed the morbidity and mortality rates of the coronavirus disease (COVID-19) pandemic in different prefectures of Japan. Under the constraint that daily maximum confirmed deaths and daily maximum cases should exceed 4 and 10, respectively, 14 prefectures were included, and cofactors affecting the morbidity and mortality rates were evaluated. In particular, the number of confirmed deaths was assessed, excluding cases of nosocomial infections and nursing home patients. The correlations between the morbidity and mortality rates and population density were statistically significant ($p$-value < 0.05). In addition, the percentage of elderly population was also found to be non-negligible. Among weather parameters, the maximum temperature and absolute humidity averaged over the duration were found to be in modest correlation with the morbidity and mortality rates. Lower morbidity and mortality rates were observed for higher temperature and absolute humidity. Multivariate linear regression considering these factors showed that the adjusted determination coefficient for the confirmed cases was 0.693 in terms of population density, elderly percentage, and maximum absolute humidity ($p$-value <0.01). These findings could be useful for intervention planning during future pandemics, including a potential second COVID-19 outbreak.

## 1. Introduction

The COVID-19 outbreak was first reported in China in 2019 [1,2] and spread worldwide in early 2020. Japan declared a state of emergency in seven (of 47) prefectures on 7 April 2020 and extended it to all prefectures on 13 April 2020. The state of emergency was withdrawn on 25 May 2020. During this state of emergency, unlike many other countries where city lockdowns were enforced, in Japan, citizens self-isolated. The mortality rate (per population) in Japan is relatively low compared to the global rate; the total number of confirmed deaths in Japan is 846 (25 May 2020), corresponding to 6.72 per million people [3]. Although a straightforward comparison is infeasible, this number is smaller than that of many other countries with the same order of magnitude of population: 541, 504, 435, and 295 in Italy, the United Kingdom, France, and the United States, respectively, but larger than 5.18 and 4.0 in Indonesia and Australia, respectively (25 July 2020).

An additional difficulty in understanding the morbidity rate is the unreliability of the diagnosis of COVID-19. The number of polymerase chain reaction (PCR) tests, a simple and cost-effective method, is limited in Japan, partly because of its reliability. Therefore, chest CT is used for a fast-track, highly accurate diagnosis [4]. Some patients do not exhibit any common pandemic symptoms [5], thereby complicating morbidity rate assessment in different areas (e.g., population composition).

Morbidity and mortality statistics have been updated every day in each prefecture in Japan, which provides a good opportunity for local studies. A notable feature of Japan is that no medical collapse has been reported. In addition, due to the health insurance system, free medical care for COVID-19 has been offered. Thus, the reliability of the mortality rate is more accurate than that of the morbidity rate, because patients with weak or mild symptoms may not be tested. However, to avoid nosocomial infections and medical resource shortages, it was suggested that people with specific symptoms (e.g., fever with temperature >37.5 °C for no more than four consecutive days) stay home and avoid seeking medical attention, unless they had been in close contact with an infected person(s) or had recently visited foreign countries. Such a policy may result in longer latency in the reported cases.

In general, coronaviruses are considered to spread mainly by respiratory droplets and contact via droplets [6]. Droplets tend to fall to the ground close to the infected host. Droplet transmission is typically limited to short distances, generally less than 2 m. There exist some hypotheses about transmission due to airborne transmission that remain in flight for one hour or longer [7]. For both mechanisms, the ambient condition potentially influences the duration of droplet and airborne spread. Several co-factors potentially influence COVID-19 morbidity/mortality rates [8–14]; among them, ambient conditions have been considered here.

The effect of ambient temperature on the mortality was discussed in Wuhan [8]. Positive and negative associations were found between daily COVID-19 death counts and daily temperature difference and absolute humidity, respectively. The effect of high temperature and humidity on the transmission of COVID-19 was discussed using relative humidity as a measure [9]. Their finding suggested that high temperature and humidity may suppress COVID transmission. Furthermore, the effect of weather on COVID-19 cases employing a case in Jakarta was presented in [10]. They reported that only average temperature is correlated with the pandemic spread. The effect of ambient temperature on the confirmed cases was discussed in more than 100 Chinese cities, and it was concluded that there is no evidence supporting that COVID-19 case counts would decline when the weather becomes warmer [12]. The effect of ambient temperature and absolute humidity on the confirmed cases was investigated in cities in China, and the researchers commented that the epidemic might gradually ease partially due to rising temperatures [13]. Instead, no correlation with UV and temperature on the transmission of COVID-19 was reported in [15].

Following Chinese studies, case studies in different countries have been reported. Briz-Redón and Serrano-Aroca [16] evaluated the spatiotemporal analysis of temperature in the cases of early COVID-19 evolution in Spain. Pirouz et al. [17] discussed the correlation between daily confirmed cases and temperature, humidity, and velocity with multivariate analysis in Italy. Application of neural networks for its estimation is also discussed in [18]. A similar attempt has been made in Oslo, Norway [19]. Recent studies have confirmed the effect of temperature and relative humidity on morbidity rates in Brazil [20,21]. From these studies, it is difficult to derive a consistent conclusion on the effect of the weather on the spread of COVID-19. Studies of influenza suggested the importance of ambient conditions for its spread: lower spread for higher humidity (e.g., [22,23]).

Studies with wider scopes included global data analysis, discussing how temperature and humidity are correlated with the infection and fatality rates of the COVID-19 pandemic [24,25]. The region of interest is wide (country level) in these studies, and thus it is not directly applicable to the ruling or regulation. In addition, some modeling studies have been proposed. However, parameter setting is not easy for this type of novel virus spread [14,26]; in most modeling studies, the parameters relating to the weather or population cannot be given explicitly. Instead, the effect of population density on the spreading effect of the epidemic has been discussed under some assumptions [27].

Nevertheless, none of the aforementioned research and modeling studies simultaneously considered the impact of population density and ambient conditions. A question that arises here is To what extent do ambient conditions and population density influence morbidity and mortality rates in different cities? Unlike the aforementioned studies, a major feature of Japan is the relative homogeneity of the health insurance and care system without medical collapse during this pandemic. In addition, the difference in household wealth is relatively small in Japan [28]. The average annual salary per population is USD 34,400 to 39,900 (USD 1 = JPY 107). The standard deviation of household consumption in each prefecture is 10% or less [28]. With all these demographic factors, the data sample discussed here provides a convenient case study with less bias. In a recent study [29], we examined the time course of the morbidity rate of different prefectures in Japan and found that the durations of the spread and decay stages can be characterized by population density, temperature, and absolute humidity. An additional factor would be the ratio of the elderly to the entire population; in Japan, this ratio reached 28.4% [30], which is ranked the highest globally.

This novel study aimed to evaluate the effect of ambient temperature and humidity on mortality and morbidity rates in different prefectures in Japan. Additionally, it considered the influence of population density and composition. To the best of the authors' knowledge, this is the first study to highlight the environmental factors' effect during COVID-19 in Japan. The model of Japan provides an interesting case study for different factors, as the medical service and social reaction is almost uniform nationwide and high-quality data were recorded properly. If the correlation of the pandemic with population density and ambient conditions is significant, the findings will be useful to set the level and duration for a strict lockdown period for each city considering the environmental factors and in planning future pandemic measures.

The organization of this study is as follows. In Section 2, the data sources of COVID-19 in Japan and weather data are mentioned. Then, the statistical method for data processing is explained briefly. In Section 3, effect of population density, elderly population, and ambient conditions on the morbidity/mortality rates are evaluated statistically. Based on the evaluation, multivariate linear regression has been conducted to estimate the morbidity/mortality rate from these parameters. In Section 4, provides discussion of the results including the limitation. The conclusion is given in Section 5.

**2. Material and Methods**

*2.1. Data Source*

In this study, we used three datasets. The first involved the confirmed daily positive cases and deaths in each prefecture [31]. This dataset is based on the report by the Ministry of Health Labor and Welfare [32]. We used time-integrated data until 25 May 2020—when the emergency state was terminated. According to the dataset, 16 prefectures had confirmed total deaths and daily positive counts higher than 4 and 10, respectively. These prefectures were defined as infected. The remaining prefectures were excluded due to a lower number of infected cases, which is simply because of the self-isolation, including the discouragement from moving to other prefectures after the declaration. In Japan, to avoid nosocomial infections and medical resource shortages, it was suggested that people with symptoms (e.g., fever >37.5 °C for no more than four consecutive days) stay home and not seek immediate medical attention unless they had been in close contact with infected people or had recently visited a foreign country. Some patients have been reported to be asymptomatic [5], making the statistical study of COVID-19 more complex. Then, the positive rate of the test varied from 2.2% to 34.8% for different prefectures. Unlike other diseases, the number of confirmed cases/deaths are counted even when patients are found to be infected after their death. For this data collection, we use the mortality rate in this study as a metric rather than the case fatality rate.

For comparison, two sets of data were prepared: (1) the number of confirmed deaths excluding and including nosocomial/nursing facility infection, and (2) the total confirmed positive cases. This is to avoid the data of cluster infection for high-risk groups, resulting in a higher possibility of death.

Note that among some of the 16 selected prefectures, as shown in Figure 1, the number of victims due to nosocomial/nursing facility infection was not always reported. Thus, such prefectures were excluded from the comparisons. Among others, the area of Hokkaido is one to two magnitudes larger than the other prefectures. Thus, several peaks in the number of cases are observed in different cities with larger distances between them than those in other adjacent prefectures. In the Gunma prefecture, three confirmed deaths occurred, except for in Isezaki City, where substantial nosocomial infections were reported (15 victims). Thus, we used data from the Gunma prefecture, excluding Isezaki City, for accurate comparison of mortality rates. In addition, Saitama was also excluded since it did not report humidity data (see below for the third dataset). Two prefectures with unclear nosocomial/nursing facility infections were also excluded.

The second dataset comprises the population and the area of the prefectures. Based on the evidence that more than 90% of the victims are older than 60 years and because the retirement age in Japan is 65, which may potentially influence morbidity rates, we set the threshold as 65. For a total of 14 prefectures and one city, the first and second datasets are listed in Table 1. Note that the rationale for choosing 25 May as a reference date is the end of state of emergency, and then, the daily confirmed death over Japan was 20 (128 million population); the daily confirmed death was smaller than 100 for one month after that.

The third dataset comprises weather data for each prefecture. They are extracted from the weather reports generated by the Japan Meteorological Agency [33]. In our previous study, we had studied the correlation between environmental conditions and the duration of the pandemic from its spread to decay periods [29]. We extended this investigation to the prefectures defined in Table 1. We then estimated the start and end dates of the spread and decay stages, as defined in Table 2. To validate the effect of ambient factors in different phases of the pandemic, we computed ambient features for three time frames: during the spreading stage $D_S$ (from $T_{SS}$ to $T_{SE}$), during the decaying stage $D_D$ (from $T_{DS}$ to $T_{DE}$), and during both stages (from $T_{SS}$ to $T_{DE}$).

To consider the mortality rate, which is affected by many factors, it should be noted that the metrics are averaged over the duration of the spread stage, decay stage, and the entire period. The duration-averaged values of temperature, absolute humidity, wind velocity, and daylight hours were calculated from the data available from the internet site mentioned above, as listed in Table 3. The latitude of Japan considered here is N 33°36′ (Fukuoka) to N 36°35′ (Ishikawa), except for Okinawa of N 26°12′), and thus, total solar radiation may be marginally influenced with this measure.

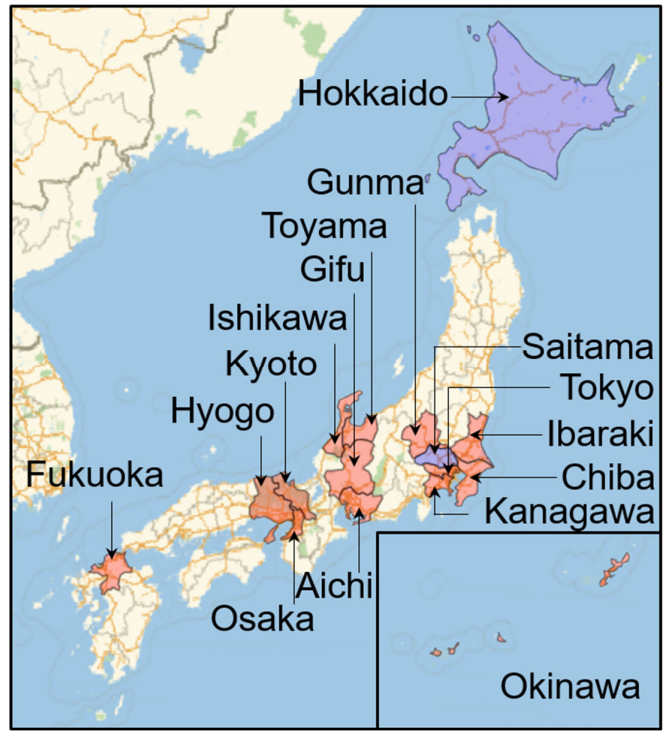

**Figure 1.** Map of Japan with prefectures included (excluded) in this study highlighted with red (blue) color.

**Table 1.** Population and population density of 14 prefectures, in addition to the percentage of the elderly population, where confirmed deaths and daily confirmed positives are larger than 4 and 10, respectively. The data of confirmed cases and deaths were counted until 25 May, 2020.

| Prefectures | Population (×1000) | Density (capita/km²) | Total Cases | Confirmed Deaths | Confirmed Deaths (Ex.) [†] | Cases /1M | Elderly (>65 years) (%) |
|---|---|---|---|---|---|---|---|
| Aichi | 7552 | 1460.0 | 507 | 34 | 16 | 67.1 | 25.1 |
| Chiba | 6259 | 1217.4 | 904 | 44 | 27 | 144.4 | 27.8 |
| Fukuoka | 5104 | 1024.8 | 672 | 25 | 20 | 131.7 | 27.9 |
| Gifu | 1987 | 187.3 | 150 | 7 | 7 | 75.5 | 30.1 |
| Gunma | 1942 | 304.6 | 149 | 19 | 19 | 76.7 | 29.9 |
| Hyogo | 5466 | 650.4 | 699 | 40 | 33 | 127.9 | 29.1 |
| Ibaraki | 2860 | 470.4 | 168 | 10 | 10 | 58.7 | 29.5 |
| Ishikawa | 1138 | 271.7 | 296 | 24 | 6 | 260.1 | 29.6 |
| Kanagawa | 9198 | 3807.5 | 1336 | 76 | 59 | 145.2 | 25.3 |
| Kyoto | 2583 | 560.1 | 358 | 15 | 15 | 138.6 | 29.2 |
| Okinawa | 1453 | 637.5 | 81 | 6 | 6 | 55.7 | 22.2 |
| Osaka | 8809 | 4631.0 | 1781 | 80 | 45 | 202.2 | 27.6 |
| Tokyo | 13,921 | 6354.8 | 5170 | 292 | 210 | 371.4 | 23.1 |
| Toyama | 1044 | 245.6 | 227 | 21 | 10 | 217.4 | 32.3 |

[†] Excluding nosocomial infection in confirmed deaths.

**Table 2.** Starting and terminating dates of the spread and decay stages of COVID-19 in different prefectures in Japan. $T_{SS}$ ($T_{DS}$) and $T_{SE}$ ($T_{DE}$) denote the start and end dates for the spread (decay) stages of the pandemic.

| Prefectures | Spread Stage | | Decay Stage | |
|---|---|---|---|---|
| | $T_{SS}$ | $T_{SE}$ | $T_{DS}$ | $T_{DE}$ |
| Aichi | 22-February | 30-March | 1-April | 27-April |
| Chiba | 19-March | 2-April | 13-April | 5-May |
| Fukuoka | 22-March | 1-April | 9-April | 27-April |
| Gifu | 25-March | 4-April | 6-April | 17-April |
| Gunma | 25-March | 5-April | 9-April | 22-April |
| Hyogo | 19-March | 4-April | 7-April | 4-May |
| Ibaraki | 16-March | 28-March | 8-April | 23-April |
| Ishikawa | 24-March | 3-April | 8-April | 8-May |
| Kanagawa | 19-March | 3-April | 11-April | 19-May |
| Kyoto | 16-March | 2-April | 5-April | 9-May |
| Okinawa | 28-March | 3-April | 10-April | 25-April |
| Osaka | 18-March | 6-April | 13-April | 6-May |
| Tokyo | 17-March | 3-April | 10-April | 7-May |
| Toyama | 1-April | 13-April | 18-April | 30-April |

**Table 3.** Duration-averaged temperature ($T$), absolute humidity ($H$), wind velocity ($V_{air}$), and daylight hours ($DL$) in each prefecture. $D_S$ and $D_D$ represent time frames during the spread and decay stages of the pandemic, respectively, as listed in Table 2. $T_{ave}$, $T_{max}$, and $T_{min}$ represent the daily average, maximum, and minimum temperatures, respectively. $H_{ave}$, $H_{max}$, and $H_{min}$ represent the daily average, maximum, and minimum absolute humidity values, respectively. $V_{air}$ represents the daily averaged wind velocity.

| | Spread Duration ($D_S$) | | | | | | | | | |
|---|---|---|---|---|---|---|---|---|---|---|
| Prefectures | $T_{ave}$ | $T_{max}$ | $T_{min}$ | $T_{diff}$ | $H_{ave}$ | $H_{max}$ | $H_{min}$ | $H_{diff}$ | $V_{air}$ | $DL$ |
| Aichi | 10.1 | 14.8 | 6.0 | 8.8 | 5.9 | 7.9 | 4.4 | 3.5 | 3.3 | 5.3 |
| Chiba | 12.4 | 16.1 | 8.1 | 8.1 | 6.6 | 9.5 | 4.6 | 4.9 | 4.5 | 4.5 |
| Fukuoka | 14.2 | 17.5 | 11.3 | 6.2 | 8.8 | 11.0 | 6.9 | 4.1 | 3.1 | 3.5 |
| Gifu | 12.0 | 16.4 | 7.7 | 8.7 | 6.7 | 8.4 | 4.9 | 3.5 | 2.7 | 4.2 |

|  | $T_{ave}$ | $T_{max}$ | $T_{min}$ | $T_{diff}$ | $H_{ave}$ | $H_{max}$ | $H_{min}$ | $H_{diff}$ | $V_{air}$ | DL |
|---|---|---|---|---|---|---|---|---|---|---|
| Gunma | 10.6 | 15.3 | 5.4 | 9.9 | 5.7 | 7.5 | 4.6 | 3.0 | 2.6 | 4.4 |
| Hyogo | 12.7 | 16.4 | 9.1 | 7.3 | 7.2 | 9.6 | 5.3 | 4.2 | 3.7 | 5.1 |
| Ibaraki | 10.3 | 17.1 | 3.4 | 13.7 | 5.7 | 8.5 | 3.7 | 4.8 | 2.7 | 7.0 |
| Ishikawa | 9.9 | 14.7 | 5.7 | 9.0 | 5.9 | 7.2 | 4.1 | 3.1 | 4.6 | 4.4 |
| Kanagawa | 12.4 | 16.7 | 8.0 | 8.7 | 6.8 | 9.7 | 4.7 | 5.0 | 4.4 | 4.6 |
| Kyoto | 11.5 | 16.6 | 6.8 | 9.8 | 6.4 | 8.6 | 4.7 | 3.9 | 2.4 | 4.6 |
| Okinawa | 21.3 | 24.0 | 18.8 | 5.1 | 14.7 | 17.4 | 12.4 | 5.0 | 4.3 | 1.9 |
| Osaka | 12.7 | 17.0 | 8.9 | 8.1 | 6.7 | 8.9 | 5.1 | 3.9 | 2.6 | 5.1 |
| Tokyo | 11.7 | 16.7 | 6.7 | 10.0 | 6.4 | 9.3 | 4.5 | 4.8 | 3.3 | 5.4 |
| Toyama | 9.7 | 14.6 | 5.2 | 9.4 | 6.3 | 7.7 | 4.7 | 3.0 | 3.2 | 4.4 |
| **Decay Duration ($D_D$)** | | | | | | | | | | |
|  | $T_{ave}$ | $T_{max}$ | $T_{min}$ | $T_{diff}$ | $H_{ave}$ | $H_{max}$ | $H_{min}$ | $H_{diff}$ | $V_{air}$ | DL |
| Aichi | 13.0 | 18.3 | 8.6 | 9.7 | 6.5 | 8.4 | 4.9 | 3.5 | 3.9 | 6.1 |
| Chiba | 15.1 | 19.1 | 11.2 | 7.8 | 8.4 | 10.3 | 6.4 | 3.9 | 4.4 | 4.6 |
| Fukuoka | 14.0 | 17.5 | 10.9 | 6.6 | 7.3 | 9.4 | 5.7 | 3.7 | 3.6 | 4.7 |
| Gifu | 12.6 | 18.2 | 7.7 | 10.6 | 5.1 | 6.6 | 3.6 | 3.0 | 3.5 | 6.5 |
| Gunma | 11.5 | 16.3 | 7.2 | 9.1 | 6.3 | 8.4 | 4.9 | 3.4 | 3.2 | 5.2 |
| Hyogo | 15.5 | 19.0 | 12.4 | 6.6 | 8.1 | 9.5 | 6.0 | 3.5 | 4.0 | 5.3 |
| Ibaraki | 10.8 | 15.6 | 6.4 | 9.1 | 6.5 | 8.2 | 4.9 | 3.3 | 3.5 | 4.7 |
| Ishikawa | 13.1 | 17.3 | 9.2 | 8.1 | 7.0 | 8.7 | 5.4 | 3.3 | 4.4 | 4.5 |
| Kanagawa | 16.6 | 20.7 | 13.0 | 7.7 | 9.8 | 11.7 | 7.7 | 4.0 | 3.9 | 4.9 |
| Kyoto | 14.7 | 20.1 | 10.0 | 10.1 | 7.1 | 9.0 | 5.3 | 3.7 | 2.5 | 5.0 |
| Okinawa | 19.8 | 22.1 | 17.6 | 4.5 | 11.8 | 14.1 | 10.0 | 4.0 | 5.0 | 3.2 |
| Osaka | 16.2 | 20.6 | 12.3 | 8.3 | 8.1 | 10.2 | 6.3 | 3.9 | 2.7 | 5.3 |
| Tokyo | 14.4 | 19.2 | 9.9 | 9.3 | 8.6 | 10.7 | 6.7 | 3.9 | 3.2 | 5.0 |
| Toyama | 12.1 | 17.6 | 7.7 | 9.9 | 7.5 | 9.1 | 5.8 | 3.3 | 3.7 | 3.9 |
| **All Duration, from $T_{SS}$ to $T_{DE}$** | | | | | | | | | | |
|  | $T_{ave}$ | $T_{max}$ | $T_{min}$ | $T_{diff}$ | $H_{ave}$ | $H_{max}$ | $H_{min}$ | $H_{diff}$ | $V_{air}$ | DL |
| Aichi | 11.3 | 16.2 | 7.1 | 6.2 | 8.2 | 4.6 | 3.5 | 11.3 | 3.5 | 5.5 |
| Chiba | 13.8 | 17.8 | 9.6 | 7.3 | 9.7 | 5.4 | 4.3 | 13.8 | 4.3 | 4.9 |
| Fukuoka | 14.9 | 19.0 | 11.4 | 8.6 | 10.8 | 6.8 | 3.9 | 14.9 | 3.3 | 5.5 |
| Gifu | 12.2 | 17.2 | 7.7 | 5.8 | 7.4 | 4.2 | 3.3 | 12.2 | 3.2 | 5.5 |
| Gunma | 11.1 | 16.0 | 6.2 | 5.9 | 7.7 | 4.6 | 3.1 | 11.1 | 2.9 | 5.2 |
| Hyogo | 14.2 | 17.8 | 10.9 | 7.5 | 9.3 | 5.5 | 3.8 | 14.2 | 3.8 | 5.6 |
| Ibaraki | 10.4 | 15.8 | 4.8 | 6.2 | 8.3 | 4.4 | 3.9 | 10.4 | 2.9 | 5.5 |
| Ishikawa | 12.1 | 16.3 | 8.0 | 6.6 | 8.1 | 4.9 | 3.2 | 12.1 | 4.2 | 4.7 |
| Kanagawa | 15.1 | 19.4 | 11.2 | 8.6 | 10.8 | 6.6 | 4.3 | 15.1 | 4.0 | 5.0 |
| Kyoto | 13.6 | 18.9 | 8.8 | 6.8 | 8.8 | 5.0 | 3.8 | 13.6 | 2.4 | 4.9 |
| Okinawa | 20.1 | 22.4 | 17.9 | 12.5 | 14.8 | 10.6 | 4.1 | 20.1 | 4.6 | 2.5 |
| Osaka | 14.3 | 18.6 | 10.4 | 7.2 | 9.3 | 5.6 | 3.7 | 14.3 | 2.6 | 5.4 |
| Tokyo | 13.3 | 18.3 | 8.5 | 7.6 | 9.9 | 5.7 | 4.2 | 13.3 | 3.2 | 5.3 |
| Toyama | 10.9 | 16.1 | 6.2 | 6.8 | 8.4 | 5.2 | 3.2 | 10.9 | 3.4 | 4.3 |

*2.2. Statistical Analysis*

A statistical study was conducted to analyze the correlation of different factors on both mortality and morbidity rates. The software JMP (SAS Institute, Cary, NC, USA) was used in this study. In order to specify dominant factors influencing the rates, *p*-value was used. We determined the pairwise correlations by calculating the Spearman's rank correlation between the number of confirmed positive cases, confirmed death cases, and different environmental and demographic parameters. Correlation matrix with partial correlation probability and CI of correlation were calculated. After that, with the same software, multivariate analysis [34] was conducted in terms of the factors. We considered linear regression for data least-squares fitting after considering multicollinearity. Statistical significance was accepted at $p < 0.05$.

## 3. Results

### 3.1. Effect of Population Density and Elderly Population

Figure 2 shows the relationship between confirmed positive cases and confirmed deaths, including and excluding nosocomial infections and nursing home patients. A modest correlation was observed between positive cases per million and population density ($R^2 = 0.394$), whereas a slight and mild correlation was observed for confirmed deaths ($R^2 = 0.097$) and excluding nosocomial infection ($R^2 = 0.259$). This result suggests that population density should be considered as a factor that implicitly represents social distancing, as is similar to our previous study that discussed the pandemic's duration [29].

When the cases and deaths for the elderly population were considered, the same tendency was observed; $R^2 = 0.363$ for cases, and $R^2 = 0.078$ and $R^2 = 0.210$ for deaths with and without nosocomial infections (not shown to avoid repetition), respectively. Instead, as shown in Figure 3, the morbidity and mortality rates normalized by population density are modestly correlated with the percentage of the elderly, especially for confirmed deaths excluding nosocomial infections ($R^2 = 0.482$). This factor is thus considered in the multivariate analysis study presented later.

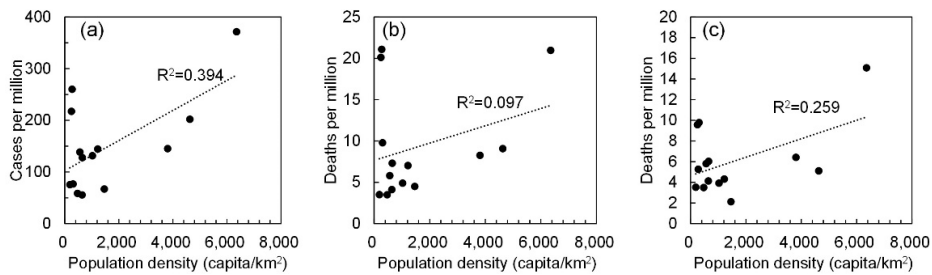

**Figure 2.** Correlation between population density and number of confirmed positive cases and fatality. The number of (**a**) positive cases, confirmed deaths (**b**) including and (**c**) excluding those caused by nosocomial infection.

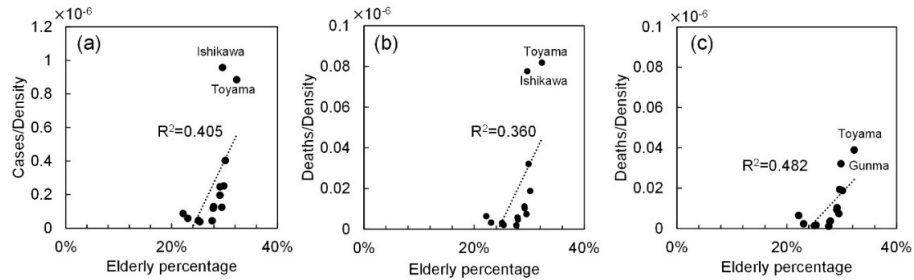

**Figure 3.** Correlation between the number of confirmed positive cases and fatality normalized by the population density and the percentage of the elderly population. The number of (**a**) positive cases, confirmed deaths (**b**) including and (**c**) excluding those caused by nosocomial infection.

### 3.2. Effect of Ambient Conditions

Several ambient factors potentially influence morbidity and mortality rates. Our study considered temperature and absolute humidity. Most previous studies reported the maximum, average, or difference (diurnal variation range) of ambient temperature (e.g., see [8] and [35]). Our study also considered the minimum temperature. Recent reports on influenza suggest the importance of absolute humidity rather than its relative value [22,23]; however, we considered the maximum, average, minimum, and difference values of absolute humidity as parameters. The daily average wind velocity and daylight hours were also considered. Regression analysis was conducted for all metrics averaged over the duration of the spread and decay stages and the total duration.

Table 4 lists the coefficients of determination for different metrics. For most parameters, the averaged values over the total stage provided the highest correlation rather than those over the other two durations. As an example, Figure 4 shows the correlation between the number of confirmed positive cases and fatality normalized by the population density and the daily maximum temperature and diurnal absolute humidity. A moderate correlation was observed among the daily maximum temperature, diurnal absolute humidity, and cases per population density. Table 5 lists the Spearman's rank correlation for different parameters. The ambient factor was normalized by population density as aforementioned. A moderate correlation was also observed with the daily maximum temperature, daily maximum, and diurnal absolute humidity and percentage of elderly population. Correlation was weak with wind velocity and daylight hours.

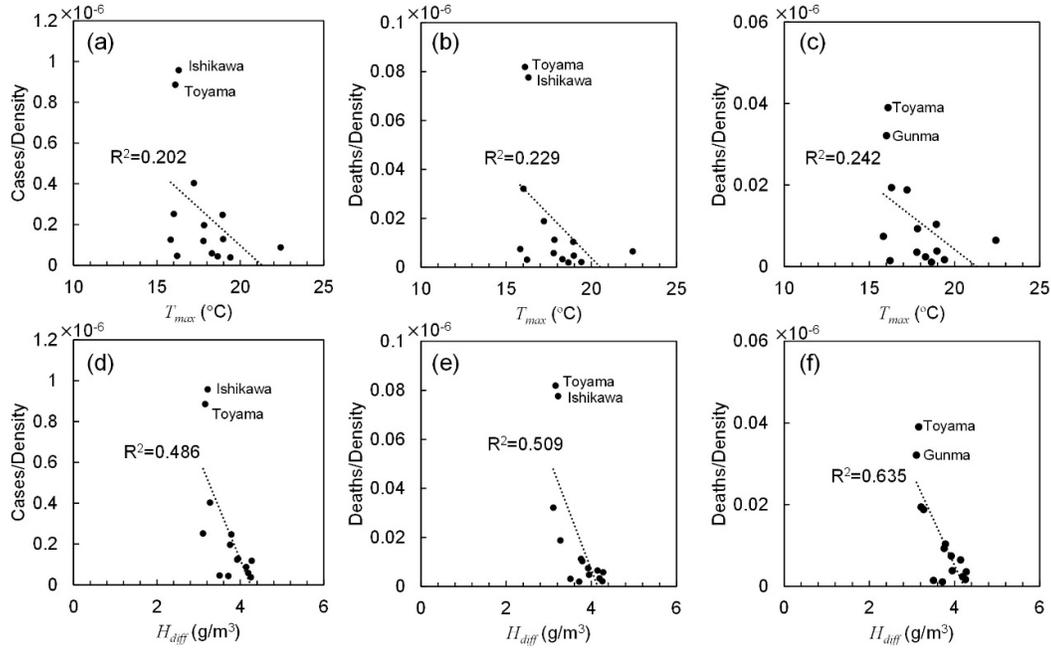

**Figure 4.** Correlation between the number of confirmed positive cases and fatality normalized by the population density and (**a–c**) the daily maximum temperature and (**d–f**) diurnal absolute humidity averaged over total duration. The number of (**a,d**) positive cases, confirmed deaths (**b,e**) including and (**c,f**) excluding those caused by nosocomial infection.

**Table 4.** Coefficient of determination for different metrics: (i) cases, (ii) death, (iii) death excluding nosocomial infection, (iv) cases normalized by density, (v) death normalized density, and (vi) death excluding nosocomial infection normalized by population density.

|  |  | (i) | (ii) | (iii) | (iv) | (v) | (vi) |
|---|---|---|---|---|---|---|---|
| Population density | | 0.393 | 0.097 | 0.259 | — | — | — |
| Elderly density | | 0.363 | 0.078 | 0.210 | 0.225 | 0.185 | 0.295 |
| Elderly percentage | | 0.009 | 0.014 | 0.007 | 0.405 | 0.360 | 0.482 |
| $T_{ave}$ | $D_S$ | 0.073 | 0.143 | 0.041 | 0.151 | 0.157 | 0.122 |
|  | $D_D$ | 0.000 | 0.035 | 0.011 | 0.164 | 0.173 | 0.274 |
|  | Total | 0.009 | 0.075 | 0.020 | 0.158 | 0.173 | 0.216 |
| $T_{max}$ | $D_S$ | 0.089 | 0.161 | 0.035 | 0.175 | 0.181 | 0.130 |
|  | $D_D$ | 0.008 | 0.019 | 0.001 | 0.143 | 0.166 | 0.229 |
|  | Total | 0.003 | 0.081 | 0.006 | 0.202 | 0.229 | 0.242 |
| $T_{min}$ | $D_S$ | 0.053 | 0.114 | 0.054 | 0.105 | 0.116 | 0.112 |
|  | $D_D$ | 0.001 | 0.041 | 0.019 | 0.147 | 0.147 | 0.246 |
|  | Total | 0.013 | 0.069 | 0.034 | 0.122 | 0.134 | 0.192 |
| $T_{diff}$ | $D_S$ | 0.007 | 0.027 | 0.047 | 0.015 | 0.021 | 0.043 |
|  | $D_D$ | 0.026 | 0.042 | 0.048 | 0.071 | 0.055 | 0.128 |
|  | Total | 0.026 | 0.042 | 0.078 | 0.036 | 0.036 | 0.101 |
| $H_{ave}$ | $D_S$ | 0.076 | 0.091 | 0.043 | 0.055 | 0.055 | 0.048 |
|  | $D_D$ | 0.017 | 0.002 | 0.019 | 0.099 | 0.061 | 0.142 |
|  | Total | 0.006 | 0.026 | 0.004 | 0.095 | 0.080 | 0.127 |
| $H_{max}$ | $D_S$ | 0.069 | 0.123 | 0.032 | 0.152 | 0.149 | 0.131 |
|  | $D_D$ | 0.016 | 0.001 | 0.019 | 0.127 | 0.081 | 0.160 |
|  | Total | 0.005 | 0.038 | 0.003 | 0.160 | 0.138 | 0.191 |
| $H_{min}$ | $D_S$ | 0.086 | 0.084 | 0.036 | 0.044 | 0.039 | 0.025 |
|  | $D_D$ | 0.011 | 0.002 | 0.016 | 0.089 | 0.051 | 0.117 |
|  | Total | 0.011 | 0.024 | 0.004 | 0.079 | 0.060 | 0.089 |
| $H_{diff}$ | $D_S$ | 0.001 | 0.107 | 0.002 | 0.463 | 0.488 | 0.546 |
|  | $D_D$ | 0.052 | 0.001 | 0.031 | 0.347 | 0.277 | 0.384 |
|  | Total | 0.006 | 0.074 | 0.000 | 0.485 | 0.509 | 0.635 |
| $V_{air}$ | $D_S$ | 0.034 | 0.058 | 0.007 | 0.020 | 0.022 | 0.044 |
|  | $D_D$ | 0.035 | 0.000 | 0.091 | 0.023 | 0.027 | 0.003 |
|  | Total | 0.001 | 0.008 | 0.032 | 0.015 | 0.017 | 0.015 |
| DL | $D_S$ | 0.023 | 0.007 | 0.012 | 0.012 | 0.010 | 0.014 |
|  | $D_D$ | 0.021 | 0.077 | 0.025 | 0.045 | 0.086 | 0.018 |
|  | Total | 0.008 | 0.007 | 0.000 | 0.035 | 0.053 | 0.029 |

**Table 5.** Spearman's rank correlation for cases normalized by density, death normalized density, and death, excluding nosocomial infection normalized by population density.

| Parameters | | Cases/Density | | Deaths/Density | | Deaths/Density (Ex.) | |
|---|---|---|---|---|---|---|---|
|  |  | $p$ | $p$-value | $p$ | $p$-value | $p$ | $p$-value |
| Elderly percentage | | 0.864 | <0.0001 | 0.824 | <0.001 | 0.842 | <0.001 |
| $T_{ave}$ | | −0.456 | 0.101 | −0.489 | 0.076 | −0.456 | 0.101 |
|  | | −0.565 | <0.05 | −0.539 | <0.05 | −0.543 | <0.005 |
|  | | −0.503 | 0.067 | −0.543 | <0.05 | −0.508 | 0.064 |
| $T_{max}$ | | −0.526 | 0.050 | −0.551 | <0.05 | −0.471 | 0.089 |
|  | | −0.631 | <0.05 | −0.574 | <0.05 | −0.560 | <0.005 |
|  | Total | −0.475 | 0.086 | −0.535 | <0.05 | −0.473 | 0.088 |
| $T_{min}$ | $D_S$ | −0.385 | 0.175 | −0.446 | 0.110 | −0.442 | 0.114 |
|  | $D_D$ | −0.524 | 0.055 | −0.506 | 0.065 | −0.511 | 0.062 |
|  | Total | −0.429 | 0.126 | −0.477 | 0.084 | −0.453 | 0.104 |
| $T_{diff}$ | $D_S$ | 0.234 | 0.422 | 0.280 | 0.333 | 0.311 | 0.280 |
|  | $D_D$ | 0.317 | 0.269 | 0.273 | 0.345 | 0.289 | 0.317 |
|  | Total | 0.315 | 0.273 | 0.326 | 0.255 | 0.375 | 0.187 |
| $H_{ave}$ | $D_S$ | −0.314 | 0.275 | −0.353 | 0.215 | −0.331 | 0.248 |
|  | $D_D$ | −0.560 | <0.05 | −0.465 | 0.094 | −0.469 | 0.091 |
|  | Total | −0.496 | 0.071 | −0.476 | 0.085 | −0.450 | 0.107 |
| $H_{max}$ | $D_S$ | −0.578 | <0.05 | −0.569 | <0.05 | −0.534 | <0.05 |
|  | $D_D$ | −0.570 | <0.05 | −0.497 | 0.070 | −0.488 | 0.076 |
|  | Total | −0.601 | <0.05 | −0.579 | <0.05 | −0.542 | <0.05 |
| $H_{min}$ | $D_S$ | −0.080 | 0.787 | −0.113 | 0.701 | −0.060 | 0.839 |
|  | $D_D$ | −0.532 | 0.050 | −0.439 | 0.116 | −0.444 | 0.112 |
|  | Total | −0.495 | 0.072 | −0.493 | 0.073 | −0.453 | 0.104 |
| $H_{diff}$ | $D_S$ | −0.665 | <0.01 | −0.583 | <0.05 | −0.579 | <0.05 |
|  | $D_D$ | −0.777 | <0.005 | −0.736 | <0.005 | −0.699 | <0.01 |
|  | Total | −0.669 | <0.01 | −0.636 | <0.05 | −0.623 | <0.05 |
| $V_{air}$ | $D_S$ | −0.160 | 0.584 | −0.081 | 0.782 | −0.187 | 0.523 |
|  | $D_D$ | −0.024 | 0.935 | 0.077 | 0.794 | −0.029 | 0.923 |
|   | Total | −0.108  | 0.714 | −0.007 | 0.982 | −0.103 | 0.725 |
| *DL* | $D_S$ | −0.464 | 0.095 | −0.411 | 0.144 | −0.446 | 0.110 |
|   | $D_D$ | −0.169 | 0.563 | −0.222 | 0.446 | −0.231 | 0.427 |
|   | Total | −0.191 | 0.513 | −0.301 | 0.296 | −0.319 | 0.267 |

*3.3. Multivariate Linear Regression*

In this subsection, the morbidity/mortality rates are estimated in terms of different factors. In Section 3.1, population density and percentage of the elderly were found to be modest, at least non-negligible factors for multivariate analysis [34]. In Section 3.2, maximum temperature and absolute humidity difference were found to be relatively important. No consistency was observed between mortality and morbidity rates. The data in Ishikawa and Toyama prefectures were considered as outliers from hierarchical clustering (see also [29]).

The difference in absolute humidity is derived from the maximum and minimum absolute humidity; at least two parameters are needed. In addition, the maximum temperature is also related to the maximum absolute humidity. In terms of variance inflation factors (VIFs), the multicollinearity was evaluated. The threshold value to differentiate small from large is generally taken as 10 [36]. From this analysis, a set of population density, elderly percentage, and absolute humidity provided estimation without multicollinearity: VIF < 3.78 for spread duration, VIF < 3.23 for decay duration, and VIF < 3.68 for total duration. Note that the maximum ambient temperature was excluded due to strong correlation with absolute humidity.

Figure 5 shows the multivariate linear regression of cases and deaths per million. Table 6 shows the determination coefficients for the three durations. As shown in Figure 5, the predicted and actual data are of good correlation with the averaged value over three stages. The highest contribution rates were the population density in the multivariate analysis (74.4%, 80.0%, and 84.5% in the cases per million, deaths per million including, and excluding nosocomial infection, respectively).

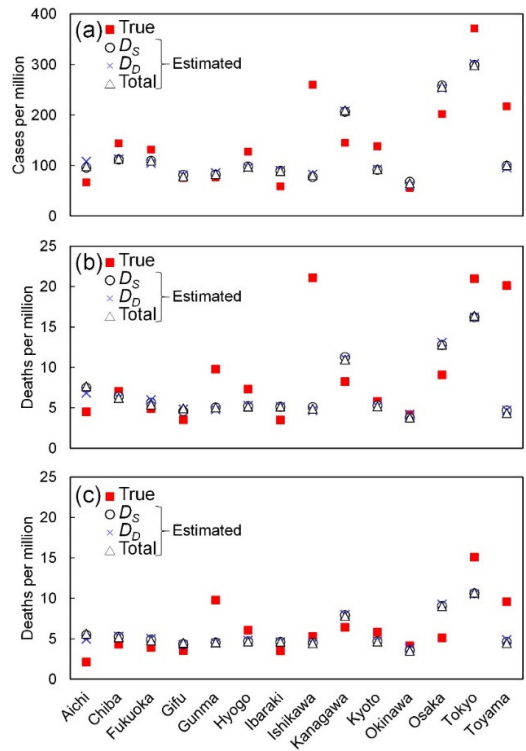

**Figure 5.** Multivariate linear regression with population density, elderly percentage, daily maximum absolute humidity averaged over spread duration. The number of (**a**) positive cases, and confirmed deaths (**b**) including and (**c**) excluding those caused by nosocomial infection.

Table 6. Coefficients of determination and adjusted $R^2$ values for multivariate linear regression.

|  | Cases | | | Deaths | | | Deaths (Ex.) [†] | | |
| --- | --- | --- | --- | --- | --- | --- | --- | --- | --- |
|  | $R^2$ | adj. $R^2$ | *p*-Value | $R^2$ | adj. $R^2$ | *p*-Value | $R^2$ | adj. $R^2$ | *p*-Value |
| $D_S$ | 0.777 | 0.693 | <0.01 | 0.659 | 0.532 | <0.05 | 0.384 | 0.153 | 0.251 |
| $D_D$ | 0.773 | 0.688 | <0.01 | 0.653 | 0.523 | <0.05 | 0.383 | 0.151 | 0.253 |
| Total | 0.776 | 0.692 | <0.01 | 0.662 | 0.536 | <0.05 | 0.386 | 0.155 | 0.249 |

[†] Excluding nosocomial infection in confirmed deaths.

## 4. Discussion

In this study, we analyzed the morbidity and mortality rates in different prefectures in Japan, where the number of confirmed deaths and daily confirmed positive counts were higher than 4 and 10, respectively. A major feature of Japan was the relative homogeneity of the health insurance and care system without medical collapse during this pandemic, in addition to household wealth. The Japanese strategy included identifying infection clusters at an early stage, to the best possible extent. However, the criteria for conducting tests (diagnosis) on potential patients may not be uniform in different prefectures; some patients may exhibit weak symptoms. Thus, after retracting the state of emergency on 25 May 2020, we processed the data for morbidity and mortality rates in 14 prefectures.

The morbidity/mortality rates were then shown to be proportional to the population density. In previous studies, this factor was not considered [12] nor was correlation between different cities considered [17]. After excluding the number of confirmed deaths in cluster infections related to hospital and care services, we observed modest correlation among different cities in terms of population density. It is worth noting that no strict closure was applied in Japan. Next, we found a good correlation between population density and the spread of COVID-19. This finding implicitly represents social distancing. In Tokyo and Osaka, which are considered among cities with the highest population densities worldwide, infection is potentially more likely to occur compared to other less dense regions. However, this may not be the case reported in other countries where strict lockdown was implemented. In Wuhan (China), the duration of the decaying stage was only 10 days, with almost no contact during the period. However, such strict lockdown may not be allowed in most countries to avoid severe social and economic damage. Therefore, this study demonstrates that population density should be considered for avoiding potential spread in future pandemics. Moreover, this finding may be useful to improve the simulation model of epidemic transmission [37,38]. The maximum temperature and absolute humidity differences were the dominant ambient factors characterizing morbidity and mortality rates. As shown in Figure 4, cases and deaths in Ishikawa and Toyama prefectures have a different tendency than that in other prefectures—as COVID-19 occurred in a very limited area in these prefectures. In general, for higher temperature and absolute humidity, the morbidity and mortality rates were decreased. For example, the population density of Hyogo (650.4 capita/km$^2$) is nearly equal to that of Okinawa (637.5 capita/km$^2$). However, the total cases in Hyogo were 8.6 times that of Okinawa. The daily maximum temperature in Hyogo was 7 °C lower than in Okinawa. This relationship can be observed in other prefectures but not all due to mild correlation with weather condition. The reason for higher correlation with absolute humidity difference is unclear. However, one potential reason would be the relatively small variation in a limited period (from mid-March to mid-May). Further study of key factors would be needed. The ambient conditions in Okinawa prefecture differ the most from those of other prefectures in Japan. If the data of Okinawa are excluded, the correlation of confirmed cases and deaths improved. In particular, the total cases and deaths normalized by population have a mild correlation with the maximum temperature and absolute humidity averaged over spread duration (from $0.13 < R^2 < 0.18$ to $0.37 < R^2 < 0.55$).

The effect of ambient conditions on the morbidity and mortality rates was shown to be modest over multiple prefecture studies. As mentioned in the introduction, this was a controversial COVID-19 issue. Our study hypothesized that this may be caused by population density, which was not considered in previous studies, as well as the uniformity of the policy, health insurance system, household wealth, etc.

The morbidity and mortality rates were roughly derived via multivariate analysis. Note that the ambient parameters are cross-correlated with each other, and thus further research and investigation are needed. Their adjusted-$R^2$ was almost the same; 0.69 ($p < 0.01$) for positive cases, and those for confirmed deaths including and excluding nosocomial infection were 0.53 ($p < 0.05$) and 0.15 ($p = 0.25$), respectively. This statistical finding may be improved for modeling studies. The correlation with the mortality rate excluding nosocomial infection was relatively low, suggesting that nosocomial infection would be a part of COVID-19 transmission at least in Japan.

Unlike previous studies that discussed the correlation with ambient condition in each city (e.g., [17]), our study explores common factors over 14 prefectures, resulting in lower $p$-value as compared to such studies. In such cases, the uncertainty of measured ambient condition would also be another factor to influence the correlation. For example, no correlation with ambient condition was observed in the analysis of 122 cities in China [12].

Note that according to the record of the Ministry of Health in Japan, no pandemic has been reported in the last 50 years [39]. Thus, a comparison with other epidemics is infeasible. Common influenza has been recorded, but only at fixed points (hospitals), making proper comparison difficult [39]. However, the finding of this study that presents the effect of population density and ambient conditions may be useful when considering measures for potential future pandemics.

**5. Conclusions**

A mild correlation was found of mortality and morbidity rates with the population density and the percentage of the elderly population, in addition to maximum absolute humidity averaged over the spread stage under Japanese policy. The multivariate linear regression provided adjusted coefficients of determination, which were 0.69 and 0.53 ($p < 0.05$) for positive cases and confirmed deaths, respectively. Our results suggested that with population and weather data, we can estimate the number of cases and deaths, at least in Japanese cities. Although the date and duration of the pandemic were different even in Japan, our estimation presented mild correlation, providing useful information for the planning of policy and medical resources. With our findings, more customized guidelines can be developed, specific to where and when different measures can be applied to restrict the adverse effects caused by a potential pandemic in the future, including a second wave of COVID-19. The limitation of this study is that the weather data in different prefectures are similar to each other due to the limited period (March to May 2020), and thus further data are needed for a general conclusion. The controversy in previous studies may be potentially caused by the population density and elderly population percentage, as those were not considered in most studies. Thus, these factors should be included for proper comparisons with the tendencies of international cities.